\documentstyle[emulateapj]{article}

\def\lae{\mathrel{<\kern-1.0em\lower0.9ex\hbox{$\sim$}}}
\def\gae{\mathrel{>\kern-1.0em\lower0.9ex\hbox{$\sim$}}}

\slugcomment{Accepted for publication in the Astrophysical Journal, November 10 1999 issue.}
 
\lefthead{C\^ot\'e et al.}
\righthead{Andromeda II}
 
\begin{document}
 
\title{Internal Kinematics of the Andromeda II Dwarf Spheroidal Galaxy}
 
\author{Patrick C\^ot\'e\altaffilmark{1,2}, Mario Mateo\altaffilmark{3}, Edward W. Olszewski\altaffilmark{4} and K.H. Cook\altaffilmark{5}}

\altaffiltext{1}{California Institute of Technology, Mail Stop 105-24, Pasadena, CA 91125}
\altaffiltext{2}{Sherman M. Fairchild Fellow}
\altaffiltext{3}{Department of Astronomy, 821 Dennison Building, University of Michigan, Ann Arbor, MI 48109}
\altaffiltext{4}{Steward Observatory, University of Arizona, Tucson, AZ 85721}
\altaffiltext{5}{Lawrence Livermore National Laboratory, Livermore, CA 94550}

 
 
\begin{abstract}
The High Resolution Echelle Spectrometer on the Keck I telescope has been used 
to measure nine radial velocities having a median precision of $\simeq$ 2 km s$^{-1}$ for
seven red giants belonging to the isolated dwarf spheroidal galaxy Andromeda II (And II).
We find a weighted mean radial velocity of 
${\overline{v}}_r$ = --188$\pm$3 km s$^{-1}$
and a central velocity dispersion of $\sigma_0 = 9.3^{+2.7}_{-2.6}$ km s$^{-1}$.
There may be evidence for a radial velocity gradient 
across the face of the galaxy, although the significance 
of this result is low due to the small number
of stars having measured velocities. Our best estimate for the global mass-to-light
ratio of And II is $M/L_V = 20.9^{+13.9}_{-10.1}M_{\odot}/L_{V,{\odot}}$.
This value is similar to those of several Galactic dwarf spheroidal galaxies and is 
consistent with the presence of a massive dark halo in And II. 
\end{abstract}
 
 
 
 
%

\section{Introduction}

The nine dwarf spheroidal (dSph) galaxies belonging to the Milky Way
are among the faintest known galaxies in the universe. Studies of their
internal kinematics began with the pioneering work of Aaronson
(1983), who reported
an unexpectedly large internal velocity dispersion of 6.5 km s$^{-1}$ for the Draco
dSph galaxy: a result based on radial velocities for three carbon stars. 
Subsequent work has not only confirmed this 
result but has established that {\sl every} Galactic dSph galaxy
has a central velocity dispersion in the range
6-13 km s$^{-1}$ ($e.g.$, Suntzeff et al. 1993; Hargreaves et al. 1994; Vogt et al. 1995;
Mateo et al. 1998). These anomalous velocity dispersions constitute
the single most compelling piece of evidence that
the Galactic dSph galaxies are embedded in massive dark halos.

Long-term monitoring of the radial velocities of red giants belonging to the Galactic
dSphs, and the acquisition and analysis of greatly expanded radial velocity samples, 
suggest that several alternative explanations for
the high velocity dispersions of these galaxies --- such as binary
stars, pulsations in the atmospheres of the most luminous dSph stars
and non-isotropic velocity distributions --- no longer appear viable 
(Olszewski, Aaronson \& Hill 1996; Hargreaves, Gilmore \& Annan 1996; 
Pryor \& Kormendy 1990). The remaining
alternatives to the conclusion that dSph galaxies are enveloped in 
dark halos of mass $\sim$ 2$\times10^7 M_{\odot}$ (Mateo et al. 1993) are 
severe heating by the Galactic tidal field (Kuhn \& Miller 1989; Klessen \& Kroupa 1998;
cf. Piatek \& Pryor 1995)
and/or the possibility that these diffuse galaxies occupy a ``low-acceleration 
regime" in which Newtonian gravity does not strictly apply (Milgrom 1983; 1990).
Although all nine Galactic dSphs have now had their internal kinematics studied 
(albeit in varying levels of detail), it is clear that some fundamental questions 
remain unanswered. 

As the nearest of the dSph galaxies associated with M31, Andromeda II (And II) 
is a promising target for a kinematic study. Since this galaxy is relatively 
isolated (van den Bergh 1972;  K\"onig et al. 1993), 
the detection of a high central velocity dispersion would
vitiate arguments that the internal motions of Galactic dSphs are
dominated by non-equilibrium dynamics. In this {\it Letter}, we present
an analysis of the internal kinematics of And II based
on nine radial velocities
for seven red giants in And II obtained with the Keck I telescope.

\section{Observations and Reductions}

\subsection{Object Selection and HIRES Spectroscopy}

Red giant members of And II were selected from the study of 
C\^ot\'e, Oke \& Cohen (1999) who present $VI$ photometry and intermediate-resolution
spectroscopy for 50 candidate red giants in And II. 
Echelle spectra for And II member giants were obtained on the nights of 14-15 and 15-16 October 
1998 using the High Resolution Echelle Spectrometer (HIRES; Vogt et al. 1994) on the 
Keck I telescope. The detector was binned 2$\times$2, giving an effective area of 
1024$\times$1024 pixels. A single readout amplifier was employed, adjusted to a gain 
setting of 2.4 $e^{-1}$/ADU. The cross-disperser and grating angles were fixed at 
$-0.05^{\circ}$ and 0.0$^{\circ}$, respectively. Since the cross disperser was used in first
order, these angles yielded a spectral coverage of 4480 $\lae \lambda\lambda \lae$
6900 \AA . The entrance aperture was limited to 1\farcs15$\times$7\farcs0 with the C5 decker.
Exposure times for all And II stars were 3600s, with Th-Ar comparison lamp
spectra taken immediately before and after each exposure. Observing conditions 
throughout the run were excellent.

The data were reduced in a manner identical to that described in Vogt et al. (1995).
A total of 29 separate apertures were 
extracted for each program object, although after some experimentation it was decided
to use only eight orders in measuring radial velocities ($i.e.$, those spanning the 
range 4848 $\lae \lambda\lambda \lae$ 5413 \AA ). Eleven high-S/N spectra for seven 
different IAU radial velocity standard stars were also obtained during the two night
run. These spectra were reduced in a similar manner to those of the And II program
stars and were used to create a master template as described in Vogt et 
al. (1995). Radial velocities for And II red giants were then derived by 
cross-correlating their spectra against that of the master template.

A critical step in deriving mass-to-light ratios from radial velocity measurements
is the proper determination of the velocity uncertainties. This is best done 
empirically, using repeat radial velocity measurements (preferably obtained 
during separate observing runs). We have combined the 34 velocities of 14 stars
given in Vogt et al. (1995) and Mateo et al. (1998) with additional velocities for
four stars (including two And II members) obtained during this run. All velocities
were obtained using identical intrumental setups and similar reduction
procedures using IRAF.\altaffilmark{6}\altaffiltext{6}{IRAF is distributed by the National Optical
Astronomy Observatories, which are operated by the Association of Universities for
Research in Astronomy, Inc., under contract to the National Science Foundation.}
We express the velocity uncertainty of each measurement as
$$\sigma_v = \alpha / (1 + R_{TD}), \eqno{(1)}$$
where $R_{TD}$ is a parameter which measures the height of the cross-correlation 
peak relative to the local noise in the cross-correlation function, and 
$\alpha$ is a constant which must be determined empirically (Tonry \& Davis 1979).
Based on our sample of repeats, we adopt $\alpha = 12.0$ which is slightly
lower than, but consistent with, the values of 14.9 and 13.4
found by Mateo et al. (1998) and Cook et al. (1999) using smaller samples.
It is, however, inconsistent with the value of 26.4 given in Vogt et al. (1995)
although the difference is due primarily to a single discrepant 
star (Leo \#23) which was discarded since it was found to contribute nearly half of
the total $\chi^2$ for the sample. We conclude that that equation 1 with $\alpha = 12.0$
closely reflects our true velocity errors, and refer the reader to C\^ot\'e et al. (1999a)
for a full discussion of the determination of $\alpha$.

The results are summarized in Table 1, which gives for each star the
ID number, $V$ magnitude, $(V-I)$ color,
projected galactocentric distance, position angle, heliocentric Julian date and
$R_{TD}$. The final two columns give the individual and mean
heliocentric radial velocities. 
Finding charts for the program stars may be found in C\^ot\'e et al. (1999b).

\section{Analysis}

\subsection{Mean Velocity, Velocity Dispersion and the Possibility of Rotation}

Using the velocities given in Table 1, we find ${\overline v}_r = -188.1\pm2.8$ km s$^{-1}$
and $\sigma_{mle} = 7.1\pm2.1$ km s$^{-1}$ using the Pryor \& Meylan (1993)
maximum-likelihood estimators for the weighted mean radial velocity and intrinsic velocity 
dispersion.  For comparison, the bi-weight estimates (Beers, Flynn \& Gebhardt 1990) for 
the systemic velocity and velocity dispersion are $v_{r,bw} = -188.5\pm3.6$ 
km s$^{-1}$ and $\sigma_{bw} = 8.3\pm1.0$ km s$^{-1}$, respectively.

The large panel of Figure 2 shows the location our seven program stars with respect 
to the center of And II. The dashed lines show the major and minor axes of the galaxy,
where we have estimated a position angle of $\theta_0$ = 160$\pm$15$^{\circ}$ for the major 
axis of the galaxy based on a visual inspection of Figure 2 of Caldwell
et al. (1992). The ellipse indicates the geometric mean King-Michie core radius 
of $r_c =1\farcm86$ (see \S 3.2),
where $e \equiv 1 - b/a = 0.3$ (Caldwell et al. 1992). The seven stars having
measured radial velocities are found in the same quadrant of the galaxy, although 
five of these objects lie within one core radius of the galaxy's center while the 
two remaining stars are located considerably further out, at radii of 4\farcm65
and 6\farcm00.

In the upper and right panels of Figure 2, we plot the heliocentric radial velocities 
for the seven stars as a function of perpendicular distance from the major and minor axes, 
respectively.
Interestingly, the two most distant stars both have radial velocities
of $v_r \gae -180$ km s$^{-1}$, whereas the velocities of the five stars near the center 
of the galaxy fall in the range $-197 \lae v_r \lae -188$ km s$^{-1}$.
This fact, along with the galaxy's rather high ellipticity, suggests
that rotation might, at least in part, be responsible for the high velocity 
dispersion measured for And II. Such a result would be surprising since 
only one Galactic dSph galaxy (Ursa Minor) is known to be rotating, and even in
this case, the rotation is dynamically unimportant: $v_{{\rm rot},0}/\sigma_0 \sim 0.5$
(Hargreaves et al. 1994; Armandroff, Olszewski \& Pryor 1995). In the next section, we 
discuss the possible effects of rotation on the derived mass-to-light ratio for And II.

\subsection{Mass-to-Light Ratio}

We have estimated the mass-to-light ratio of And II in two different ways. First,
we have fit an isotropic, single-mass King-Michie model to the $V$-band surface brightness 
profile given in Caldwell et al. (1992), asssuming $E(B-V) = 0.062$ mag,
$A_V = 3.1E(B-V)$ and $D = 660$ kpc (C\^ot\'e et al. 1999b). 
The parameters of this model are given in Table 2, along with other fundamental
parameters for And II. Both the radii used in the fit and those quoted
in this table are geometric mean radii.
The tabulated (1$\sigma$) 
uncertainties in these parameters have been determined by fitting King-Michie models
to each of 100 simulated data sets generated from the original model and using the 
uncertainties in the observed surface brightness profile 
(see Fischer et al. 1992).
Integrating the isotropic model over all radii gives a total luminosity of
$L_V = (2.95^{+0.73}_{-0.50})\times10^6 L_{V,{\odot}}$ for And II.
Based on the seven radial velocities given in Table 1, we find a scale
velocity of $v_s = 10.8^{+3.1}_{-3.0}$ km s$^{-1}$ and a systemic velocity of
$v_0 = -185.4\pm$3.5 km s$^{-1}$ for this model. 
These values correspond to a central velocity dispersion of $\sigma_0 = 9.3^{+2.7}_{-2.6}$ km s$^{-1}$
and $M/L_V = 20.9^{+13.9}_{-10.1}M_{\odot}/L_{V,{\odot}}$,
which we adopt as our best estimate for the global mass-to-light ratio for And II.
Note that if the sample is limited to just the five stars with $r < r_c$, this
estimate drops to $M/L_V = 2.0^{+1.9}_{-1.3}M_{\odot}/L_{V,{\odot}}$, 
demonstrating that the high mass-to-light ratio found here is due to the large
velocity residuals of the two outermost stars.

As a check on the above mass-to-light ratio, we have also used
the tensor virial theorem to estimate the mass of And II, explicitly including
the possible effects of rotation.
Naturally, with a sample of only seven radial velocities, 
the exact form of the rotation law (if any) is hopelessly under-constrained. Our 
goal here is simply to explore the possible effects of
ordered motions on the inferred velocity dispersion and, hence, on the
derived mass-to-light ratio. In the right and upper panels of Figure 2 we show
the rotation-corrected velocities for the seven And II members assuming
that, over the region spanned by these stars, the galaxy is rotating as a
solid-body about the minor and major axes, respectively.
The virial mass is then given by
$$M = {2{\overline{r}}\over{G}}(v^2_{{\rm rot},0} + 3\sigma^2_{0}) \eqno{(2)}$$
where $v_{{\rm rot},0}$ is the mass-weighted rotation velocity, $\sigma_{0}$ is the mass-weighted
velocity dispersion and ${\overline{r}}$ is the harmonic radius ($e.g.$, see Meylan \& Mayor 
1986). The rotation-corrected velocities ($i.e.$, the open circles in Figure 2)
are then used to derive $\sigma_{0}$, approximating the density profile by the isotropic
King model discussed above (which assumes that mass traces light).
In the two cases of solid-body rotation about the major and
minor axes, the derived mass-to-lights ratios are 
$M/L_V = 63$ and $19M_{\odot}/L_{V,{\odot}}$, respectively. If rotation is
neglected ($i.e.$, the uncorrected velocities are used), then the virial mass 
estimator gives $M/L_V = 22$, in agreement with the preceding results.

\subsection{Discussion}

We now discuss the implications of the mass-to-light ratio of And II presented
here, concentrating on three possible interpretations: (1) Modified Newtonian Dynamics
(MOND); (2) non-equilibrium dynamics; and (3) dark matter.

\subsubsection{Modified Newtonian Dynamics}

Milgrom (1983, 1995) has argued that dSph galaxies occupy a ``low-acceleration regime"
in which Newton's second law deviates from the standard $r^{-2}$ law.
For isotropic velocity dispersion tensors, the MOND virial mass is given by
$$M = {81 \over 4} \sigma^4_{0} {(Ga_0)}^{-1} \eqno{(3)}$$
where $G$ is the gravitational constant, $\sigma_0$ is the central line-of-sight 
velocity dispersion and $a_0 = 1.2\times10^{-8}$ cm s$^{-1}$ is the MOND acceleration
constant. Adopting $\sigma_0$ = 9.3$^{+2.7}_{-2.6}$ km s$^{-1}$ (the central velocity dispersion of
the isotropic King-Michie model) and $L_V$ = (2.95$^{+0.73}_{-0.50}$)$\times10^6L_{V,{\odot}}$
gives $M/L_V = 3.2^{+4.8}_{-2.3} M_{\odot}/L_{V,{\odot}}$ for the MOND mass-to-light 
ratio, which is consistent with the mass-to-light
ratios of Galactic globular clusters. We conclude that MOND
provides an adequate description of the internal kinematics of And II, without
requiring the presence of a dark halo.

\subsubsection{Non-Equilibrium Dynamics: Tides}

One possible explanation for the high central velocity dispersions of dSph galaxies
is that these objects are not in dynamical
equilibrium as a result of time-dependent oscillations caused by the tidal field of the host 
galaxy ($e.g.$, Kuhn \& Miller 1989; Cuddeford \& Miller 1990) or outright tidal 
disruption (Kroupa 1997; Klessen \& Kroupa 1998). Might this be the case for
And II? Following Mateo et al. (1993), we calculate for each Local Group dE/dSph 
galaxy a ``stability index'', $X = \log_{10}(\rho_{V,0}/\rho_{\rm Gal})$, where 
$\rho_{V,0} \sim \mu_{V,0}/2r_c$ is the central density of the dSph galaxy 
(determined from the observed surface brightness profile and an assumed mass-to-light
ratio of $M/L_V = 2M_{\odot}/L_{V,{\odot}}$) and 
$\rho_{\rm Gal}$ is the mean density of the host galaxy interior to the position of
each satellite. The latter values have been calculated assuming logarithmic 
potentials for M31 and the Milky Way having 
circular velocities of $v_c = 260$ and $220$ km s$^{-1}$, respectively.
The results of this exercise are shown in the upper panel of Figure 3. For
And II, we find $\rho_{V,0} = 0.012\pm0.003M_{\odot}$ pc$^{-3}$ and $X$ $\sim$ 2.1,
meaning that the internal baryonic mass density in And II is $\sim$ 120
times greater than the average enclosed mass density ($i.e.$, dark matter and baryons).
Evidently, tides are unlikely to play an important role in driving the internal 
kinematics of And II.\altaffilmark{7}\altaffiltext{7}{Although M33 is 
$\sim$ 2$\times$ nearer (in projection) to And II, its influence is 
negligible compared to that of M31 due to its much lower mass ($e.g.$, $v_c \sim 80$ km
s$^{-1}$ according to Figure 3 of Zaritsky, Elston \&  Hill 1989).}

\subsubsection{Dark Matter}

The lower panel of Figure 3 shows the global mass-to-light ratios of Local Group 
dE/dSph galaxies (data from Mateo 1998) plotted as a function of their absolute magnitude. 
Clearly, And II is consistent with the established
trend between mass-to-light ratio and integrated luminosity (Kormendy 1987). In addition, the
mass-to-light ratio found here is consistent, at the 1.2$\sigma$ level, with the
suggestion (Mateo et al. 1993; Mateo 1998) that all dE/dSph galaxies studied to date
consist of luminous components having $M/L_V = 2M_{\odot}/L_{V,{\odot}}$ which are
embedded in dark matter halos of mass $M \sim 2\times10^7M_{\odot}$, as indicated by the 
dashed line in Figure 3.

\acknowledgments
 
The authors thank Phil Fischer for useful discussions, and the referee,
Tad Pryor, for his prompt and helpful comments.
PC acknowledges support provided by the Sherman M. Fairchild Foundation.
MM was partially supported by grants from the NSF during the course of this
research. EWO was partially supported by NSF grants AST 92-23967 and AST
96-19524.

\begin{deluxetable}{lcccccccc}
\tablecolumns{9}
\tablewidth{0pc}                        
\tablecaption{Radial Velocities for Red Giants in Andromeda II\label{tbl-2}}                    
\tablehead{                        
\colhead{ID} &                       
\colhead{$V$} &                      
\colhead{$(V-I)$} &                       
\colhead{$R$} &                       
\colhead{$\theta$} &
\colhead{HJD} &                      
\colhead{$R_{TD}$} &                       
\colhead{$v_r$} &                       
\colhead{${\overline{v}_r}$} \nl              
\colhead{} &                      
\colhead{(mag)} &
\colhead{(mag)} &
\colhead{($'$)} &     
\colhead{(deg)} &
\colhead{2440000+} &                      
\colhead{} &                       
\colhead{(km s$^{-1}$)} &
\colhead{(km s$^{-1}$)} 
}                        
\startdata                        
And II-5  & 21.89 & 1.72 & 1.36  & 243 & 11101.8579 & 3.97 & --192.3$\pm$2.4 & --192.6$\pm$1.6 \nl
          &       &      &       &     & 11101.9021 & 4.45 & --192.8$\pm$2.2 &                 \nl
And II-32 & 21.82 & 1.81 & 0.72  & 221 & 11101.9499 & 5.05 & --188.1$\pm$2.0 & --188.1$\pm$2.0 \nl
And II-22 & 21.65 & 1.69 & 4.65  & 195 & 11101.9961 & 3.41 & --180.1$\pm$2.7 & --178.7$\pm$1.8 \nl
          &       &      &       &     & 11102.0407 & 4.05 & --177.6$\pm$2.4 &                 \nl
And II-4  & 21.91 & 1.62 & 0.53  & 250 & 11102.7910 & 7.08 & --188.9$\pm$1.5 & --188.9$\pm$1.5 \nl
And II-11 & 21.94 & 1.63 & 1.61  & 200 & 11102.8367 & 4.97 & --194.8$\pm$2.0 & --194.8$\pm$2.0 \nl
And II-36 & 22.05 & 1.78 & 1.29  & 181 & 11102.9153 & 4.12 & --197.3$\pm$2.3 & --197.3$\pm$2.3 \nl
And II-26 & 21.82 & 1.66 & 6.00  & 191 & 11103.0083 & 4.11 & --176.4$\pm$2.3 & --176.4$\pm$2.3 \nl
\enddata                        
\clearpage

\end{deluxetable}                        

\clearpage

\begin{deluxetable}{lcrrr}
\tablecolumns{5}
\tablewidth{0pc}
\tablecaption{Observed and Derived Parameters for Andromeda II\label{tbl-2}}
\tablehead{
\colhead{Quantity} &
\colhead{Symbol} &
\colhead{Value} &
\colhead{Units} &
\colhead{Reference} 
}
\startdata
~~~Distance                  & $D$       &  660$^{+100}_{-85}$  & kpc & 4\nl
~~~True Distance Modulus     & $(m-M)_0$ &24.1$\pm$0.3  & mag & 4\nl
~~~Reddening                 & $E(B-V)$  &0.062$\pm$0.010 & mag & 3\nl
~~~Absolute Magnitude        & $M_V$     &$-11.40^{+0.23}_{-0.19}$& mag& 2,5\nl
~~~Core Radius               & $r_c$     &1.89$^{+0.47}_{-0.37}$ & arcmin& 2,5\nl
                             &           &362$^{+91}_{-71}$ & pc& 2,5\nl
~~~Half-Mass Radius          & $r_h$     &3.27$^{+1.12}_{-0.50}$ & arcmin& 2,5\nl
                             &           &627$^{+216}_{-96}$ & pc& 2,5\nl
~~~Tidal Radius              & $r_t$     &13.8$^{+11.4}_{-4.5}$  &arcmin& 2,5\nl
                             &           &2.65$^{+2.19}_{-0.86}$ & kpc& 2,5\nl
~~~Concentration             & $c$       &$0.87^{+0.37}_{-0.25}$& & 2,5\nl
~~~Ellipticity               & $e$       &0.3 & & 2\nl
~~~Position Angle            & $\theta_0$&160$\pm$15 & deg& 2,5\nl
~~~Central Surface Brightness& $\mu_{V,0}$& 24.75$\pm$0.12 & mag arcsec$^{-2}$& 2,5\nl
                             &            &  4.43$^{+0.50}_{-0.47}$ & $L_{V,{\odot}}$ pc$^{-2}$& 2,5\nl
~~~Integrated Luminosity     & $L_V$      &(2.95$^{+0.73}_{-0.50})\times10^6$&$L_{V,{\odot}}$& 2,5\nl

~~~Maximum Likelihood Mean Velocity       & ${v}_0$ & $-188.1\pm2.8$ & km s$^{-1}$& 5\nl
~~~Biweight Central Velocity              & ${v}_{r,bw}$ & $-188.5\pm3.6$ & km s$^{-1}$& 5\nl
~~~Maximum Likelihood Dispersion          & ${\sigma}_{mle}$ & 7.1$\pm$2.1 & km s$^{-1}$& 5\nl
~~~Biweight Velocity Dispersion           & $\sigma_{bw}$ &8.3$\pm$1.0 & km s$^{-1}$ & 5\nl
{\it King-Michie Model}                   & & & \nl
~~~Scale Velocity                         & $v_s$ &10.8$^{+3.1}_{-3.0}$ & km s$^{-1}$& 5\nl
~~~Central Velocity Dispersion            & $\sigma_{0}$ &9.3$^{+2.7}_{-2.6}$ & km s$^{-1}$& 5\nl
~~~Mass-to-Light Ratio                    & $M/L_V$& 20.9$^{+13.9}_{-10.1}$ & $M_{\odot}/L_{V,{\odot}}$ & 5\nl
~~~Mass-to-Light Ratio ($r < r_c$, $N_* = 5$) & & 2.0$^{+1.9}_{-1.3}$ & $M_{\odot}/L_{V,{\odot}}$ & 5\nl
{\it Virial Theorem}                         & & & \nl
~~~Mass-to-Light Ratio (no rotation)         &       & 22 & $M_{\odot}/L_{V,{\odot}}$ & 5\nl
~~~Mass-to-Light Ratio (minor axis rotation)\tablenotemark{a} &       & 19 & $M_{\odot}/L_{V,{\odot}}$ & 5\nl
~~~Mass-to-Light Ratio (major axis rotation)\tablenotemark{b} &       & 63 & $M_{\odot}/L_{V,{\odot}}$ & 5\nl
\enddata

\tablenotetext{ }{References for Table 2: 
(1) K\"onig et al. (1993);
(2) Caldwell et al. (1992);
(3) Schlegel, Finkbeiner \& Davis (1998);
(4) C\^ot\'e, Oke \& Cohen (1999b);
(5) This paper}

\tablenotetext{a}{Mass-to-light ratio found using radial velocities corrected for solid-body
rotation around the galaxy's minor axis ($e.g.$, see the right panel of Figure 2).}
\tablenotetext{b}{Mass-to-light ratio found using radial velocities corrected for solid-body
rotation around the galaxy's major axis ($e.g.$, see the upper panel of Figure 2).}

\end{deluxetable}
\clearpage

\clearpage
 
\plotone{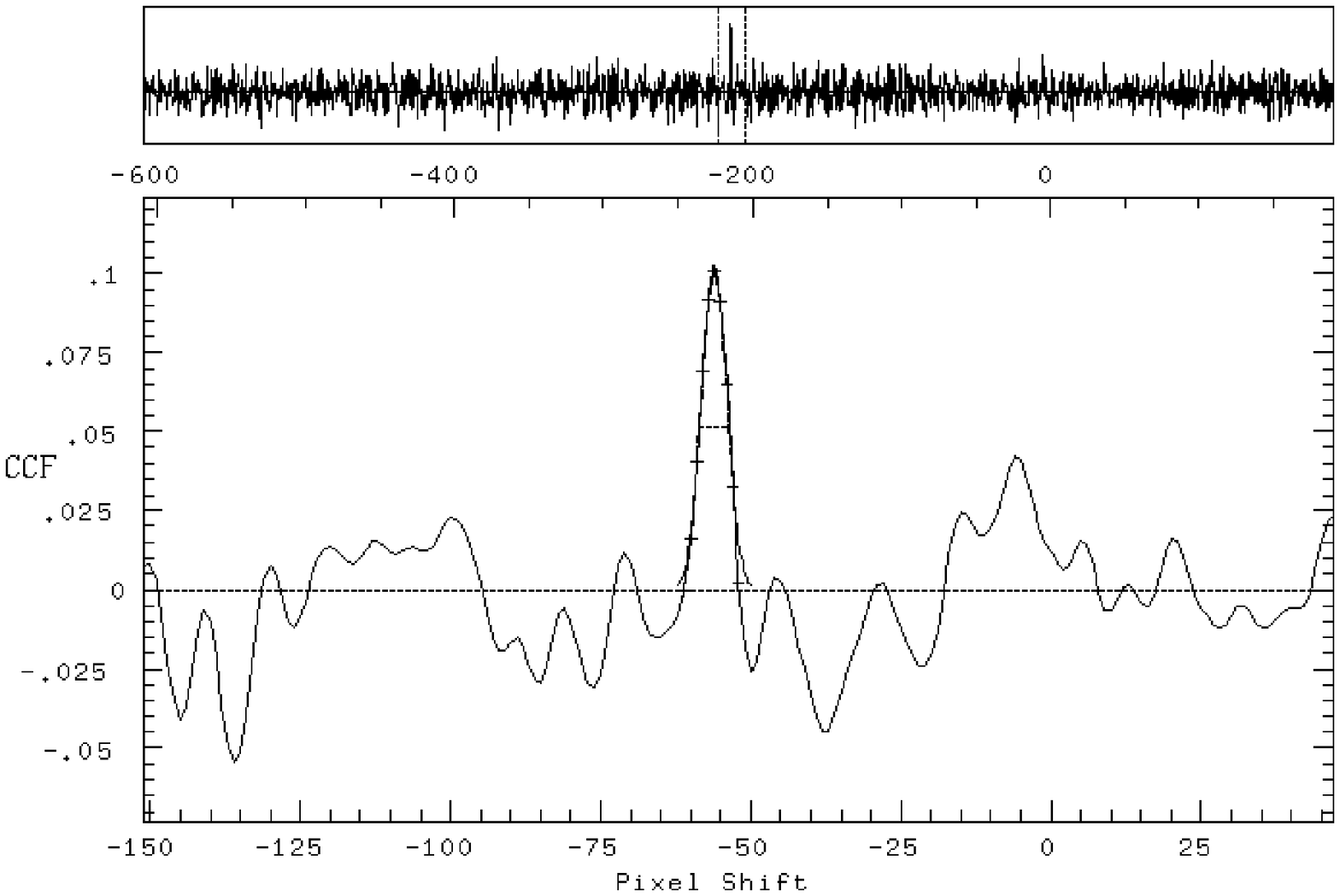}

\figcaption[and2a.01.ps]{Cross-correlation function for Andromeda II-5 obtained using
our master radial velocity template. The Tonry-Davis $R_{TD}$ value for this
cross correlation is $R_{TD} = 4.45$, slightly lower than the average of 
${\overline{R}}_{TD}$ = 4.58 for the nine measurements given in Table 1.
\label{fig1}}


\plotone{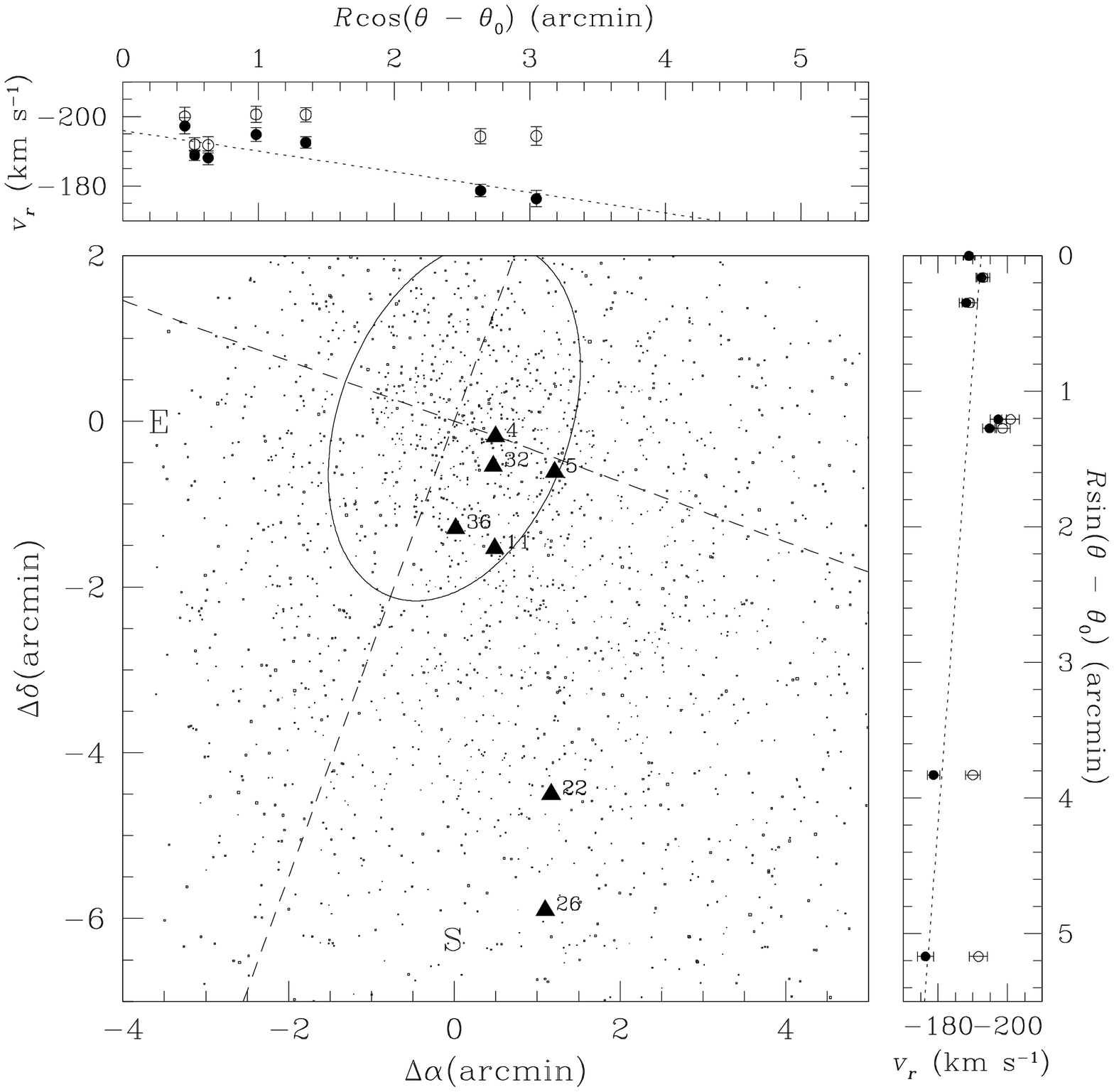}

\figcaption[and2a.02.ps]{(Large Panel) Distribution of unresolved objects in the
vicinity of Andromeda II. Red giants having one or more measured radial velocities are 
indicated by the large triangles. The core radius of Andromeda II is indicated by 
the ellipse, which has $r_c = 1\farcm89$, $a = 2\farcm26$ and $b/a = 0.7$ 
(Caldwell et al. 1992). The dashed lines indicate the minor and major axes of
the galaxy. (Upper Panel) Radial velocities for Andromeda II members plotted
against distance along the photometric minor axis (filled circles). The dotted-line
indicates the weighted least-squares fit to the points. The open circles show the
rotation-corrected velocities, assuming solid-body rotation about the {\it major} axis.
(Right Panel) Same as previous panel, except for the case of solid-body rotation 
about the photometric {\it minor} axis.
\label{fig2}}


\plotone{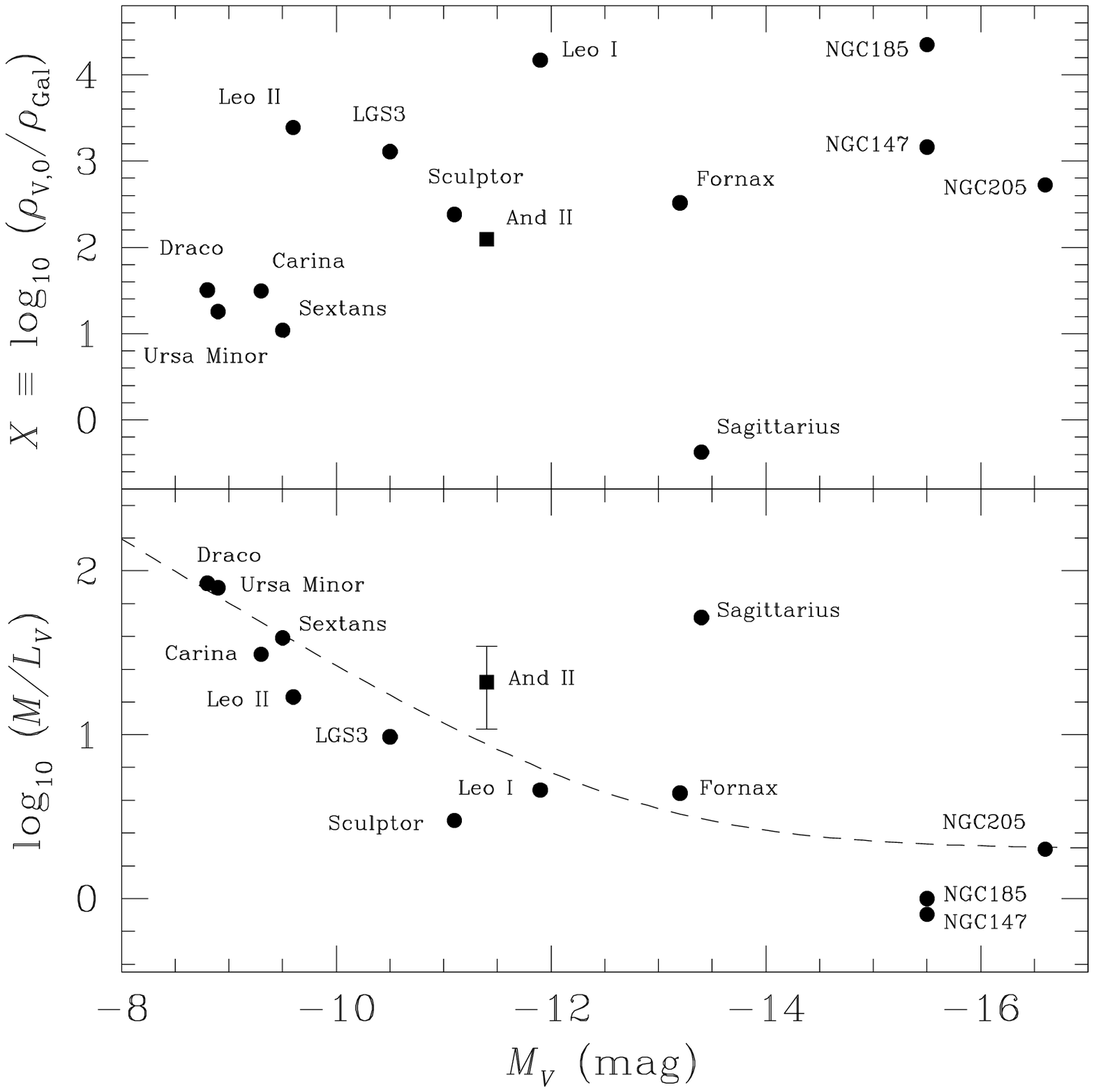}

\figcaption[and2a.03.ps]{
(Upper Panel) Stability index, $X$, for Local Group
dE/dSph galaxies plotted against $V$-band absolute magnitude (circles).
The total of the Milky Way is assumed to be $1\times10^{12}M_{\odot}$. The mass 
of M31 is taken to be twice this value.
(Lower Panel) Global $V$-band mass-to-light ratio for Local Group dE/dSph 
galaxies, plotted against $V$-band absolute magnitude (circles). The dashed 
line indicates the expected relation for dwarfs consisting of luminous 
components having $M/L_V = 2M_{\odot}/L_{V,{\odot}}$ which are embedded 
in dark matter halos of mass $M \sim 2\times10^7M_{\odot}$.
\label{fig3}}

 
\end{document}